\documentclass[fleqn,usenatbib]{mnras}

\usepackage{newtxtext,newtxmath}

\usepackage[T1]{fontenc}

\DeclareRobustCommand{\VAN}[3]{#2}
\let\VANthebibliography\thebibliography
\def\thebibliography{\DeclareRobustCommand{\VAN}[3]{##3}\VANthebibliography}

\usepackage{graphicx}	
\usepackage{amsmath}	
\usepackage{amssymb}	
\usepackage{xcolor}
\usepackage[export]{adjustbox}
\usepackage{flushend}
\usepackage{lastpage}
\usepackage[utf8]{inputenc}
\usepackage[normalem]{ulem}
\usepackage{comment}
\usepackage{url}

\usepackage[compact]{titlesec}  
\titlespacing{\section}{0pt}{12pt}{7pt}
\titlespacing{\subsection}{0pt}{9pt}{4pt}

\newcommand{\dmu}{pc\,cm$^{-3}$}	
\newcommand{\zu}{km s$^{-1}$ kpc$^{-0.5}$}

\author[Main et al.]{R.~A.~Main$^1\thanks{Email: \href{mailto:ramain@mpifr-bonn.mpg.de}{ramain@mpifr-bonn.mpg.de}}$, 
A.~Parthasarathy$^{1}$,
S. Johnston$^{2}$, 
A. Karastergiou$^{3,4}$,
A.~Basu$^{5}$,
A.~D.~Cameron$^{6,7}$, \newauthor
M.~J.~Keith$^{5}$,
L. S. Oswald$^{3,8}$,
B. Posselt$^{3,9}$,
D. J. Reardon$^{6,7}$,
X. Song$^{5}$, 
P. Weltevrede$^{5}$ \\
{$^{1}$ Max-Planck-Institut f\"{u}r Radioastronomie, Auf dem H\"{u}gel 69, D-53121 Bonn, Germany}\\
{$^{2}$ CSIRO Astronomy and Space Science, Australia Telescope National Facility, PO~Box~76, Epping NSW~1710, Australia}\\
{$^{3}$ Department of Astrophysics, University of Oxford, Denys Wilkinson Building, Keble Road, Oxford OX1 3RH, UK}\\
{$^{4}$ Department of Physics and Electronics, Rhodes University, PO Box 94, Grahamstown 6140, South Africa}\\
{$^{5}$ Jodrell Bank Centre for Astrophysics, Department of Physics and Astronomy, University of Manchester, Manchester M13 9PL, UK}\\
{$^{6}$ Centre for Astrophysics and Supercomputing, Swinburne University of Technology, P.O. Box 218, Hawthorn, Victoria 3122, Australia} \\
{$^{7}$ Australian Research Council Centre of Excellence for Gravitational Wave Discovery (OzGrav)} \\
{$^{8}$ Magdalen College, University of Oxford, Oxford OX1 4AU, UK}\\
{$^{9}$ Department of Astronomy \& Astrophysics, Pennsylvania State University, 525 Davey Lab, 16802 University Park, PA, USA}
}

\title[Scintillation Arcs with the TPA]{The Thousand-Pulsar-Array programme on MeerKAT - X. \\ Scintillation arcs of 107 pulsars}

\date{Accepted XXX. Received YYY; in original form ZZZ}

\pubyear{2022}

\begin{document}
\label{firstpage}
\pagerange{\pageref{firstpage}--\pageref{lastpage}}
\maketitle

\begin{abstract}
We present the detection of 107 pulsars with interstellar scintillation arcs at 856--1712\,MHz, observed with the MeerKAT Thousand Pulsar Array Programme.  Scintillation arcs appear to be ubiquitous in clean, high S/N observations, 
their detection mainly limited by short observing durations and coarse frequency channel resolution.
This led the survey to be sensitive to nearby, lightly scattered pulsars with high effective velocity -- from a large proper motion, a screen nearby the pulsar, or a screen near the Earth. 
We measure the arc curvatures in all of our sources, which can be used to give an estimate of screen distances in pulsars with known proper motion, or an estimate of the proper motion.  The short scintillation timescale in J1731$-$4744 implies a scattering screen within 12\,pc of the source, strongly suggesting the association between this pulsar and the supernova remnant RCW 114.  We measure multiple parabolic arcs of 5 pulsars, all of which are weakly scintillating with high proper motion.  Additionally, several sources show 
hints of inverted arclets suggesting scattering from anisotropic screens.
Building on this work, further targeted MeerKAT observations of many of these pulsars will improve understanding of our local scattering environment and the origins of scintillation; annual scintillation curves would lead to robust screen distance measurements, and the evolution of arclets in time and frequency can constrain models of scintillation.

\end{abstract}


\section{Introduction}

Observations of the radio emission of pulsars exhibits scintillation, an interference pattern owing to electron density fluctuations in the ionized interstellar medium (IISM).  Recent studies of several pulsars have revealed ``scintillation arcs'' - a parabolic distribution of power in the 2D power spectrum of scintillation commonly referred to as the secondary spectrum.  Discovered by \citet{stinebring+01}, scintillation arcs have proven a powerful probe of the distribution and geometry of scattering structures in the IISM, as well as a probe of pulsar emission regions (e.g. \citealt{pen+14}), and binary orbits (e.g. \citealt{reardon+20}).

In addition to a main parabola, many sources show `arclets', inverted parabolae of the same curvature stemming from apices on the main parabola, which persist over timescales of months.  This is difficult to explain with a picture of turbulent, isotropic, volume filling IISM, instead suggesting compact scattering structures along highly anisotropic thin `screens' \citep{walker+04, cordes+06}. Some scattering screens have been seen to persist over many years \citep{reardon+20, mckee+22, mall+22}, suggesting the extent of these screens to be $\gtrsim 1000$\,AU.  Moreover, arclets are seen to track between observations and across frequency \citep{hill+05, sprenger+22}, consistent with compact scattering structures at fixed angular positions.

It is currently unknown precisely how and where these screens originate, although screen associations have placed screens in supernova remnants \citep{cordes+04, yao+21}, HII regions \citep{mall+22}, potentially near the edge of the local bubble \citep{reardon+20, sprenger+22}, or in ionized gas surrounding hot stars \citep{walker+17}.  The electron densities required to scatter radio light at such large angles are difficult to reconcile with pressure balance in the IISM, leading to models of scattering in thin over- or under-dense plasma sheets seen edge on \citep{penlevin14}, or in magnetic filaments \citep{gwinn19}.

While most studies of scintillation arcs to date have focused on specific sources of interest, wider surveys are useful in determining the behaviour and occurrence of scintillation arcs, as well as the distribution of screens in the Milky Way. Recent surveys suggest scintillation arcs are common, if not prevalent. \citet{stinebring+22} found scintillation arcs in 19 out of 22 detected pulsars in a survey of bright, low Dispersion Measure (DM) pulsars using the Green Bank Telescope and the Arecibo Observatory. \citet{wu+22} detect scintillation arcs in 9 out of 31 sources with LOFAR at $110-190\,$MHz; scintillation becomes finer in both time and frequency, requiring very fine channelization and time sampling, resulting in harsh signal-to-noise limitations in detecting arcs.  

The Thousand Pulsar Array (TPA), is part of the Large Survey Project ‘MeerTime’ on the MeerKAT telescope \citep{bailes+20}, which has observed $1270$ pulsars in the southern sky in a wide band from 856$-$1712\,MHz (details of the TPA in \citealt{johnston+20}).  With the wide band, and the massive upgrade in sensitivity of MeerKAT compared to Parkes, the TPA can be used as a representative survey of scintillation properties of pulsars in the southern sky.

In this paper, we present the first results of scintillation using the TPA, focusing on detection and measurements of scintillation arcs of 107 pulsars.  Section \ref{sec:background} revisits the necessary theory of scintillation arcs, Section \ref{sec:data} describes the data, the reduction, the sample and curvature measurements, Section \ref{sec:results} describes our results, and Section \ref{sec:discussion} includes ramifications of our results and prospects for future work.

\section{Background}
\label{sec:background}

In this section, we briefly review the relevant theory of scintillation arcs, and refer the reader to \citet{walker+04, cordes+06} for more detailed treatment.  Additionally, in section \ref{sec:approx}, we outline limiting cases, and the approximations necessary to estimate the screen distance or pulsar proper motion from a single measure of a scintillation arc curvature.
\subsection{Scintillation Arcs}

The dynamic spectrum $I(\nu, t) = |E(\nu, t)|^{2}$ expresses the pulse intensity as a function of time and frequency.  The 2D power spectrum, or secondary spectrum $|I(\tau, f_{D})|^2$ expresses the intensity in terms of its conjugate variables $f_{D}$ and $\tau$ which are related to the Doppler shift and phase delay.
In the limit of strong scattering, the secondary resulting interference pattern can be described through the stationary phase approximation as a discrete set of deflected images.  The secondary spectrum then contains the interference of all pairs of images
\begin{align}
f_{D, ij} &= \frac{ (\boldsymbol{\theta_{i}} - \boldsymbol{\theta_{j}} ) \cdot \boldsymbol{v}_{\rm eff} }{\lambda}, \\
\tau_{ij} &= \frac{d_{\rm eff} (\theta_{i}^{2} - \theta_{j}^{2})}{2 c},
\end{align}
where $\lambda$ is the observing wavelength, and the effective distance $d_\mathrm{eff}$ and effective velocity
$\boldsymbol{v}_\mathrm{eff}$ are defined as
\begin{align}
d_{\rm eff} &= (1/s - 1) d_{p},  \\
\boldsymbol{v}_{\rm eff} &= (1/s - 1) \boldsymbol{v}_{p} + \boldsymbol{v}_{\oplus} - \boldsymbol{v}_{l} / s, \\
s &\equiv 1 - d_{l} / d_{p},
\end{align}
where $d_\mathrm{p}$, $d_\mathrm{l}$ and $\boldsymbol{v}_{\rm p}$, $\boldsymbol{v}_\mathrm{l}$ are the distances and velocities of the pulsar and the scattering screen, respectively, and $\boldsymbol{v}_{\oplus}$ is the Earth's velocity.
In weak scattering, the main pulsar image is much brighter than subimages, resulting in a sharp forward parabola.  With the main pulsar image at $\boldsymbol{\theta_{j} = 0}$, the common dependence of $\tau$ and $f_{D}$ on $\theta$ results in a relation of 
\begin{equation}
    \tau = \eta f_{D}^{2}, \quad \eta = \frac{\lambda^2 d_{\rm eff}}{2c v_{\rm eff, ||}^2}.
\end{equation}
In strong scattering from a highly anisotropic screen, subimages are of comparable brightness, and the cross-terms result in inverted parabolae (`arclets') of the same curvature with apices located on the main arc. 

Following \citet{mall+22} and \citet{sprenger+22}, we choose not to work directly with $\eta$, but rather with
\begin{equation}
W \equiv \frac{|v_{\rm eff, ||}| }{ \sqrt{d_{\rm eff} } } = \frac{\lambda}{\sqrt{2c \eta}},
\end{equation}
separating the measurements and constants from physical quantities of interest, where $W$ has the added benefit of approaching $0$ rather than diverging when $v_{\rm eff, ||} \rightarrow 0$.

\subsection{Approximations}
\label{sec:approx}

For an isolated pulsar, $W$ contains the degenerate combination of 6 unknown parameters; the pulsar's distance $d_p$ and 2D velocity $\boldsymbol{v_p}$, the screen's distance $d_l$, and either the parallel velocity $v_{l,||}$ and axis of anisotropy $\psi$ for a 1D screen, or the 2D velocity $\boldsymbol{v_{l}}$ for an isotropic 2D screen.  Even in pulsars where the proper motion and distance are independently well measured, a single measure of $W$ is insufficient to measure the screen properties, where this degeneracy is typically either broken by using annual variations of $W$ or $\eta$ \citep{reardon+20, mall+22, mckee+22, sprenger+22}, or through VLBI or multi-telescope dynamic spectra \citep{brisken+10, simard+19b}.  However, one does not typically have annual scintillation curves, or scintillation cross-spectra; under certain limits reviewed here, one can obtain meaningful information from $W$.

\textbf{Screen is near to the pulsar or the Earth -- }
A value $W$ much greater than $v_{p}$ is likely to imply local scattering, either near to the pulsar or the Earth.
When scattering is in the pulsar's local environment (e.g. supernova remnants \citealt{cordes+04, yao+21}, or eclipsing binaries, \citet{main+18, lin+21} ), then $d_{l}\rightarrow d_{p}$, and $s\rightarrow0$.  The Earth's velocity is negligible, and the only relevant terms are the distance and velocity between the source and the screen. Assuming the screen velocity is negligible, and the screen is isotropic or aligned with $\boldsymbol{v}_{\rm eff}$,
\begin{equation}
    d_{pl} \approx \frac{v_p^2}{W^2},
\label{eq:screendistpsr}
\end{equation}
where $d_{pl} \equiv d_{p}-d_{l}$.

Interstellar screens $\lesssim 5\,$pc of the Earth have been observed from scintillation arcs of B1133+16 \citep{mckee+22}, and scintillation of active galactic nuclei \citep{wang+21}.
Under the same assumptions as above, for a local screen,
\begin{equation}
    d_{l} \approx \frac{v_{\earth}^2}{W^2}.
\label{eq:screendistearth}
\end{equation}
We caution however that screen velocities or misalignment can greatly influence the estimates of $d_l$ or $d_{pl}$, as their contributions to either are squared.

\textbf{Screen is halfway to the pulsar - }
when the pulsar proper motion is unknown, it can be estimated given an assumption of the screen distance.  If e.g. the screen is halfway to the pulsar, then $d_{\rm eff} = d_{p}$, $s=1/2$, and
\begin{equation}
\mu \approx \frac{W}{\sqrt{d_p}}.
\end{equation}
While highly uncertain, this can result in a rough estimate of pulsar proper motions.  This argument is equivalent to determining proper motions using scintillation velocities, e.g. \citet{cordes+98}.

\section{Data}
\label{sec:data}

The data and processing for this paper are identical to those used in the census of \citet{posselt+22}; all of the data were taken under the Thousand Pulsar Array project \citep{johnston+20} as part of MeerTime \citep{bailes+20}.
In brief the observations were taken with the MeerKAT L-band receiver (centered at $\approx$\,1283.6\,MHz), and are folded with 8\,s integrations and $\approx0.836\,$MHz channels, corresponding to a total range of 62.5\,mHz in $f_{D}$ and $\approx0.6\mu$s in $\tau$.  
For one pulsar (see Section \ref{sec:J1744}) we used the data available at a high time resolution of 38.28\,$\mu$s to extend the range of $f_D$.
As observation duration is one of the limiting factors to detecting scintillation arcs, we limited this study to the single longest observation of every pulsar, typically $\sim 5-20$\,min, although in a few cases much longer.

\subsection{Creating Dynamic Spectra}

To create the raw dynamic spectra, radio frequency interference (RFI) is flagged and masked by the standard deviation in the off-pulse region, using methods identical to \citet{main+20, mall+22}.  After RFI masking, the off-pulse is subtracted in every time and frequency subintegration. The background-subtracted folded spectrum is then multiplied by the frequency averaged pulse profile, and summed over phase and polarization to form $I(t, \nu)$.

\begin{figure*}
\centering
\includegraphics[width=0.238\textwidth, trim=0.0cm 0cm 0cm 0.0cm, clip=true,]{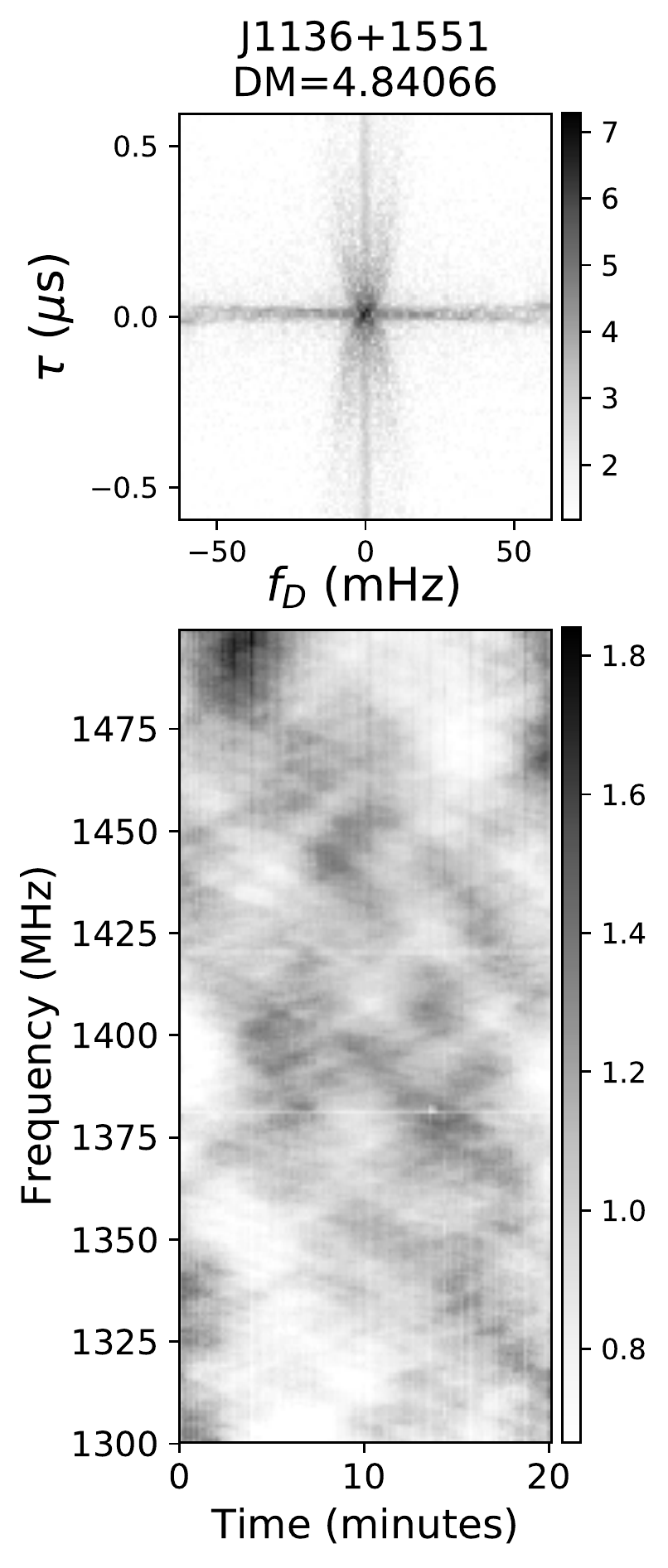}  
\includegraphics[width=0.17\textwidth, trim=2.0cm 0cm 0cm 0.0cm, clip=true]{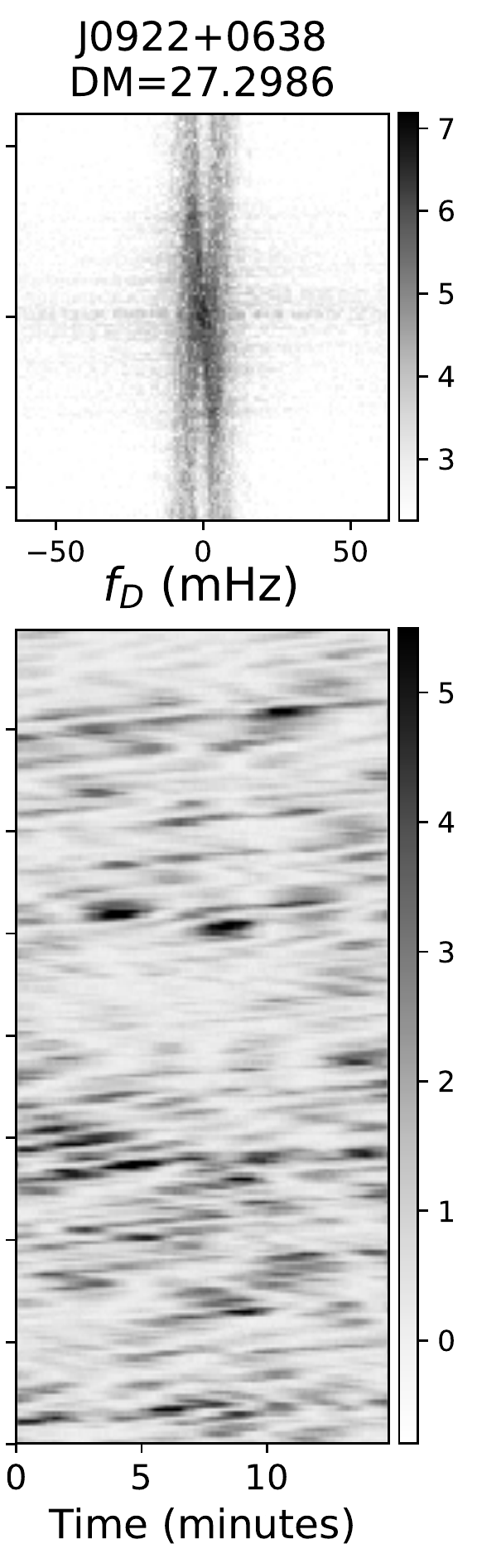} 
\includegraphics[width=0.1785\textwidth, trim=2.0cm 0cm 0cm 0.0cm, clip=true]{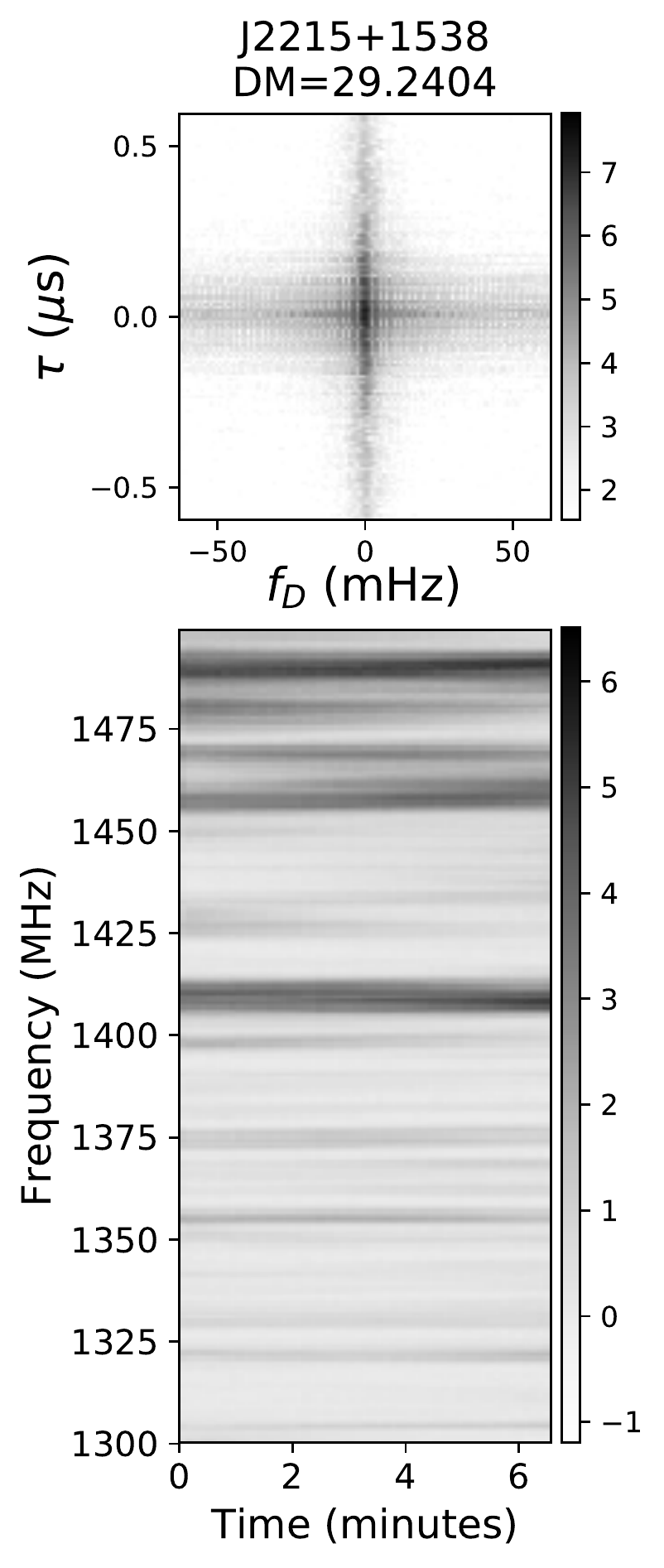} 
\includegraphics[width=0.18\textwidth, trim=2.0cm 0cm 0cm 0.0cm, clip=true]{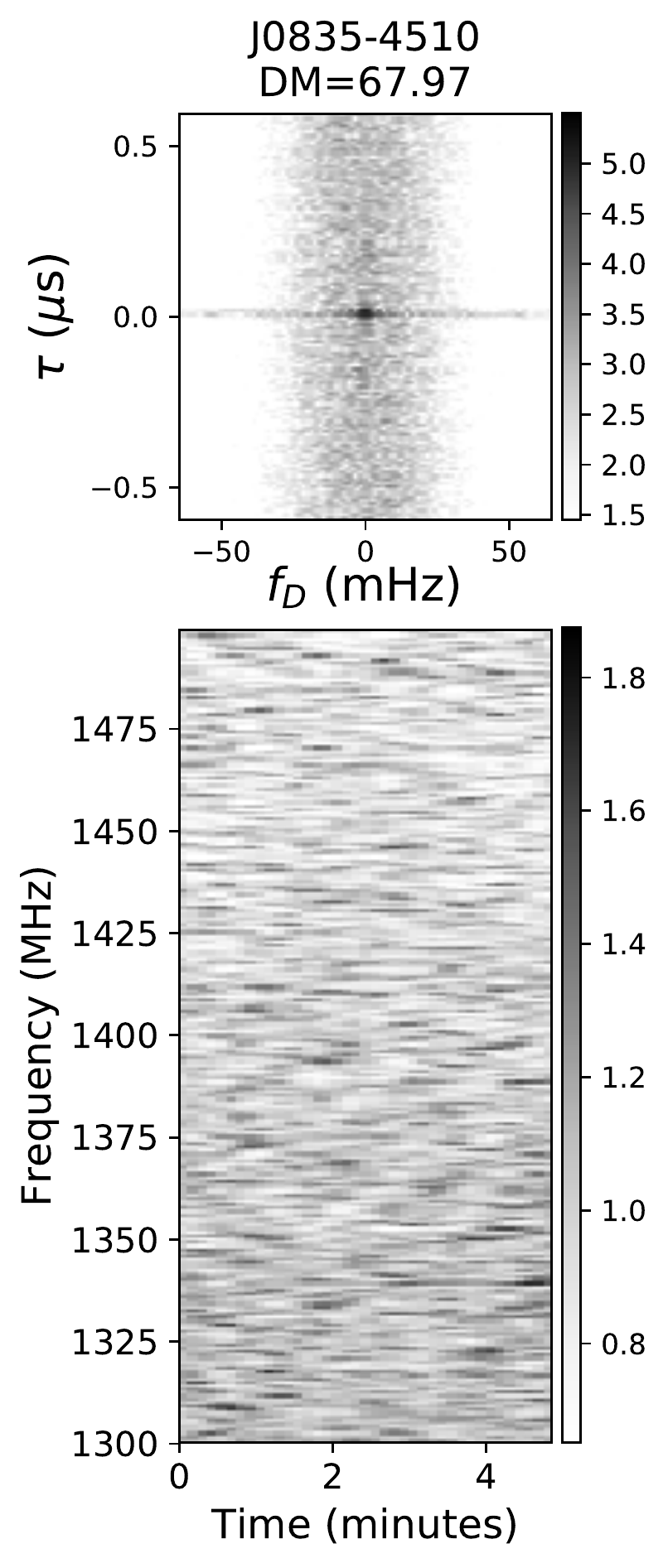} 
\includegraphics[width=0.18\textwidth, trim=2.0cm 0cm 0cm 0.0cm, clip=true]{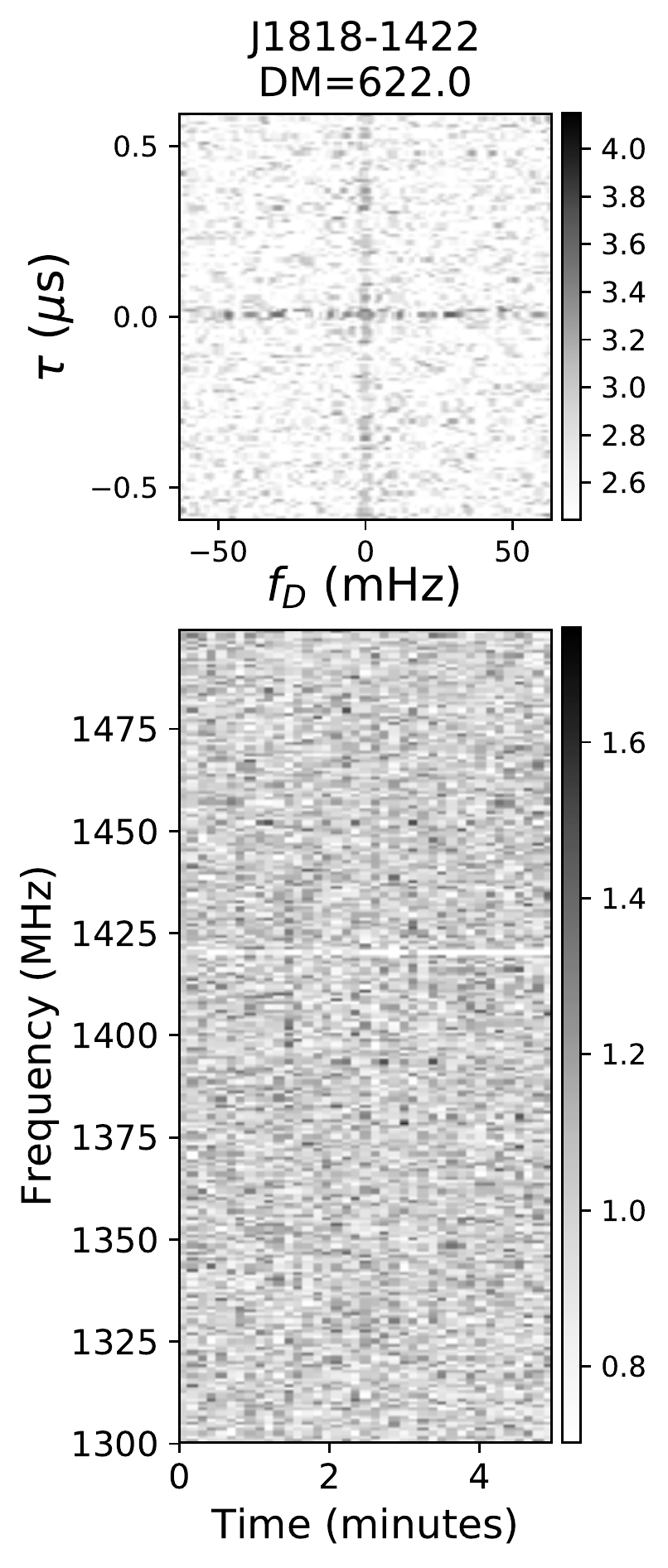} \\ 
\caption{Characteristic examples of dynamic spectra at 1.4\,GHz (\textit{bottom}) and corresponding secondary spectra (\textit{top}).  Dynamic spectra are normalized such that their average intensity is 1, and secondary spectra are plotted in log-scaling, spanning the full range in $f_D$ and $\tau$. J1136+1551 and J0922+0638 both show arcs; J1136+1551 is weakly scintillating with the intensity slightly modulated around 1, while J0922+0638 strongly scintillates with intensity extending to many times the mean.  J2215+1538 shows strong scintillation resolved in frequency, but unresolved in time due to low $v_{\rm eff}/\sqrt{d_{\rm eff}}$ and short duration. J0835$-$4510 shows strong scintillation resolved in time but far unresolved in frequency; due to high S/N, scintillation is still apparent but with greatly reduced modulation.  J1818$-$1422 is scattered to $\approx 12$\,ms at L-band \protect{\citep{oswald+21}}, corresponding to $\nu_{s} \approx 10\,$Hz, entirely unresolved by our channels. }
\label{fig:spectraexamples}
\end{figure*}

Due to the high sensitivity and wide band of MeerKAT, the dominant source of noise in the dynamic spectrum of many pulsars is not receiver noise, but rather the pulse-to-pulse variations and the window function caused by the RFI mask.  The measured dynamic spectrum is $I_{\rm meas}(t, \nu) = I_{\rm s}(t, \nu)M_{\rm RFI}(t, \nu) I_{\rm int}(t, \nu)$.  The intrinsic variations are often removed by dividing the frequency-averaged intensity from every subintegration, but this has the effect of up-weighting noisy integrations (and vice-versa), and diverges when the pulsar nulls.  We try to solve this problem using a Wiener Filter, following the same steps as \citet{lin+21}. 
The one addition is that we multiply our window function by the frequency-averaged flux in each time bin, $I_{\rm int}(t, \nu)$, in an attempt to filter out the intrinsic pulse variations. For computational reasons, the full band is split into 8 sub-bands, and the time axis is split in longer observations (> 24\,minutes, 180 time bins), and the Wiener Filter computed separately in each. This leads to slight artefacts at the boundaries, resulting in low $f_D$ noise in the secondary spectrum.  Characteristic examples of filtered dynamic spectra are shown in Figure \ref{fig:spectraexamples}.

\subsection{Creating Secondary Spectra}

Due to the $\lambda^2$ scaling of the arc curvature $\eta$, a regular 2D FFT over a wide band will result in a smearing of scintillation arcs.  We account for this by using the `NuT' transform applied in \citet{sprenger+2021}, where the Fourier Transform in time is performed on a scaled axis $t^{'} = \frac{\nu}{\nu_{\rm ref}} t$, which fixes arcs to constant curvature.  In addition, compact features of fixed angular positions on the sky will be seen at constant positions across frequency in the secondary spectrum.  We choose $\nu_{\rm ref} = 1400\,$MHz for all plots in this paper.  Before taking the Fourier transform, the mean of the dynamic spectrum was subtracted, and a $10\%$ border in both frequency and time were tapered with a Hanning window.

\subsection{Arc Detections}

For every pulsar, diagnostic plots were created which included the dynamic and secondary spectra, and 2D autocorrelation functions (ACFs) across the full band, as well as zoom ins to the two clean bands of 920-1020\,MHz, and 1300-1500\,MHz.  Any sources with observable modulation were deemed sources with either fully or partially resolved scintillation - for each of these pulsars, secondary spectra were produced with a range of $f_{D}$ limits in both clean bands.  These plots were inspected manually;  scintillation arcs were identified as sources with significant cross-power in the secondary spectrum; ie. power away from the $f_{D}=0$, $\tau=0$ axes.  Characteristic examples showing arc detections, as well as sources with partially resolved scintillation (in either time or frequency) but without an arc, are shown in Figure \ref{fig:spectraexamples}.

\subsection{Arc Curvature Fitting}
\label{sec:arcfitting}

To fit for $W$, we used the `Normalized Secondary Spectra', which remaps the $f_{D}$ axis to $f_{D,\rm norm} = f_{D} \sqrt{\tau_{\rm ref}/\tau}$ \citep{reardon+20}.  This transformation maps parabolae to vertical lines of $f_{D,\rm norm} \propto W$, and in this paper we take $\tau_{\rm ref} = 0.6\mu$s, the maximum value of $\tau$ allowed by the channel widths.  Summing over the $\tau$ axis results in the power as a function of arc curvature $S(W)$, where arcs can be identified as peaks.  We fit peaks in $S(W)$ with an inverted parabola,  where the starting values and fitting range of $W$ are chosen to vary by source, using an interactive fitter. 
Arc asymmetries may originate from local phase gradients across the scattering screen \citep{cordes+06, rickett+14}, so we fit $W$ separately for positive and negative values of $f_{D}$; final values of $W$ are the mean from both sides, while we also record the difference $\Delta W_{lr}$. 
The error of $W$ is computed from the half width at half maximum (HWHM) divided by the S/N of the peak, where the N is the standard deviation of $S(|W| > 2W_{\rm fit})$.
For sources with screen distances discussed in the text (all of which are cases of large $W$ suggesting local scattering), the two screen solutions are estimated from equations \ref{eq:screendistpsr} and \ref{eq:screendistearth}.  The screen solution near the pulsar uses previously measured proper motions, while the solution near Earth uses \textsc{scintools} \citep{scintools} to predict Earth's velocity vector in the plane of the sky in the direction of the pulsar.

\section{Results}
\label{sec:results}

Of the 1270 pulsars observed, we found 107 to have observable scintillation arcs.  The measured values of $W$, $\Delta W_{lr}$ and derived estimates of 
$V_{\rm ISS}$ are in Tables \ref{table:arcmeas} and \ref{table:arcmeasnov},
and secondary spectra plots are shown in Figure \ref{fig:SecspecPano}.

The detection of arcs is greatly limited by the short durations of observations and the coarse channels; many sources showed a clear scintillation pattern in frequency, but with too short a duration to resolve the pattern in time, and many more sources have scintles far smaller than the channel bandwidth, and are averaged out in each channel.  The channel resolution means we are biased towards detecting arcs in mildly scattered, primarily nearby / low DM sources. The short observing durations mean that only high $W$ sources will show an arc; ie. sources with large velocities, or screen near to the pulsar or Earth. This is seen in our results $-$ the inferred proper motion velocities are $\gtrsim 100\,$km/s in most sources, in many cases much larger.

\begin{figure}
\centering
\includegraphics[width=1.0\columnwidth, trim=0.0cm 0.0cm 0cm 0cm, clip=true,]{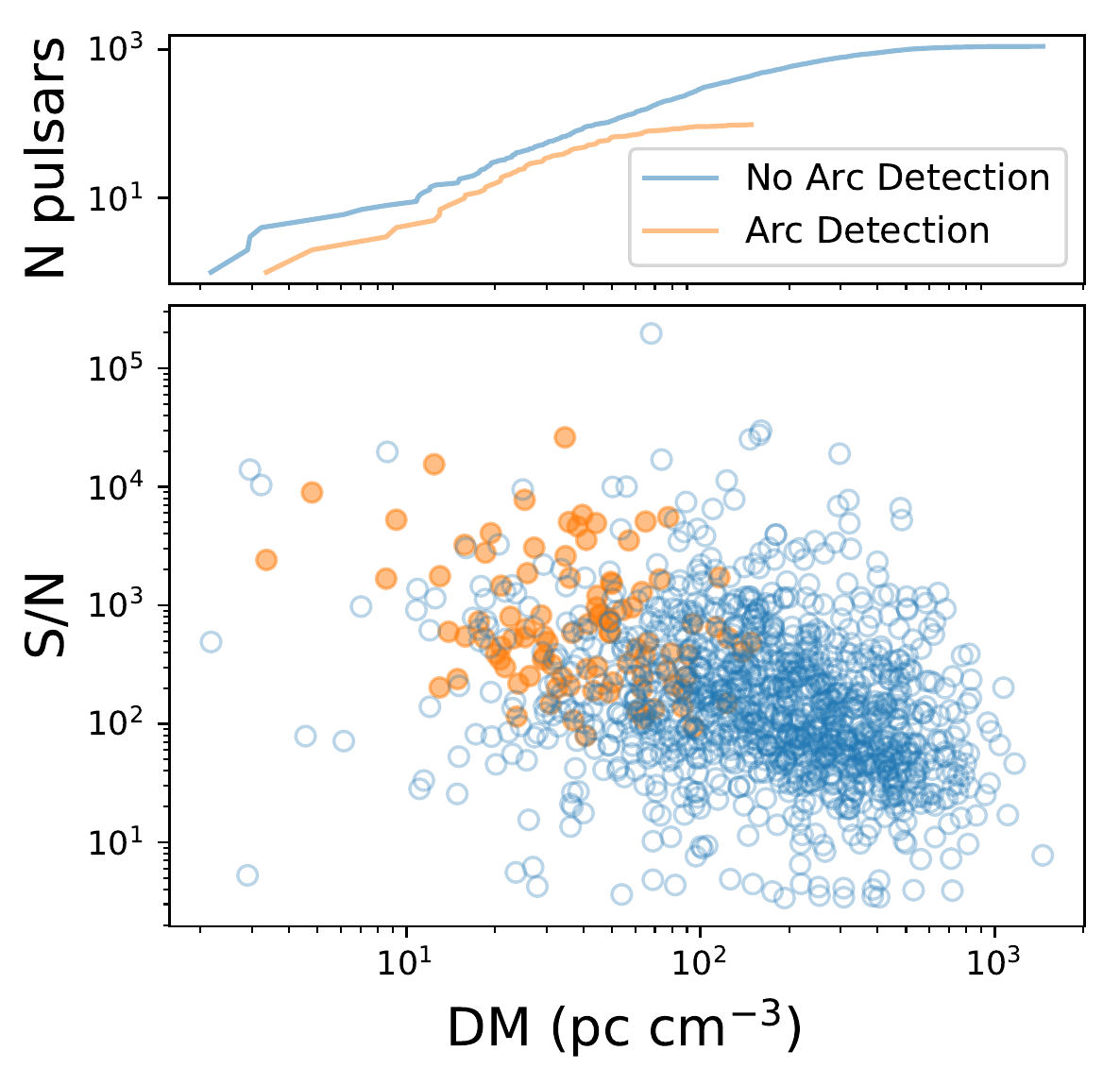} 
\vspace{-5mm}
\caption{Sample DM, S/N, and the detections of arcs.  Top: cumulative distribution of sources vs. DM.  Bottom: Scatter plot of source S/N vs. DM in the longest TPA observation for each source, which was inspected for arcs. Sources with detected arcs were found with DM$\lesssim 100\,$\dmu, S/N$>100$.}
\label{fig:Sample}
\end{figure}

In addition to the constraints from channel bandwidth and duration, the detection of arcs can be limited by S/N, or by unmasked RFI, pulsar moding or nulling, which lead to convolved noise in the secondary spectrum.  While we tried to account for the pulsar's variable emission in the dynamic spectra using a Wiener Filter, this is highly suboptimal when pulsars are nulling for $\gtrsim 50\%$ of the time, and the RFI flagging and excision may not be perfect in all observations. 
We plot the DM and S/N of our arc detections compared to the full sample in Fig. \ref{fig:Sample}.  
The sources with arcs are primarily with DM\,$\lesssim 100\,$\dmu, and with S/N $\gtrsim 100$, where higher DM sources are typically more scattered, with scintillation unresolved due to our coarse channels.

\subsection{Multiple Screens, and Evolution with Frequency}
\label{sec:multiscreen}
\begin{figure*}
\includegraphics[width=1.0\textwidth,trim=0.0cm 0.2cm 0cm 0.0cm, clip=true]{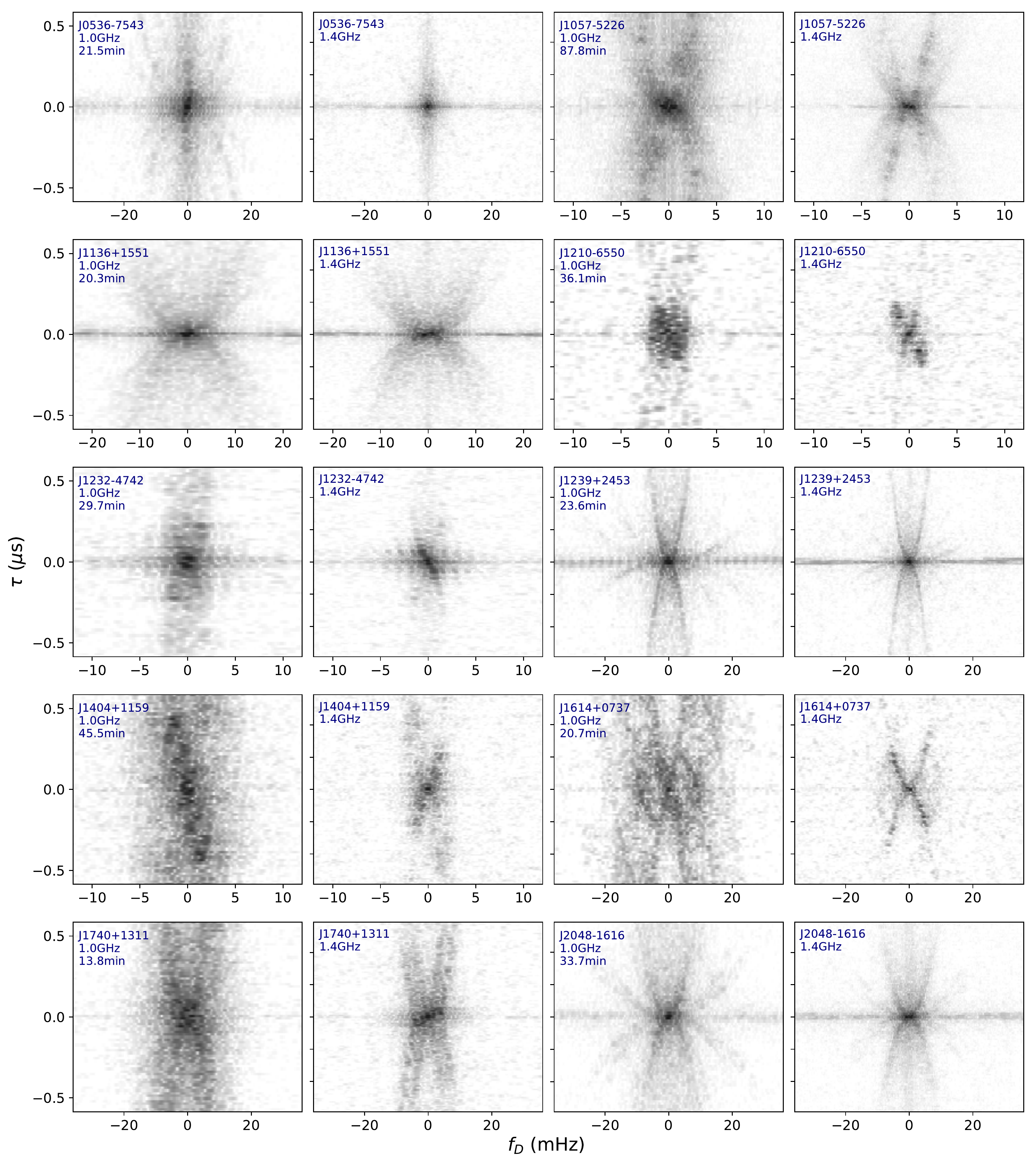}
\vspace{-7mm}
\caption{Secondary spectra of multi-arc systems, and a selection of sources showing interesting evolution across frequency, discussed in sections \ref{sec:multiscreen} and \ref{sec:evolution}. }
\label{fig:SecspecMultiscreen}
\end{figure*}

We detect 5 sources with clear evidence of multiple arcs, shown in Figure \ref{fig:SecspecMultiscreen}.  These sources all tend to be very low DM, nearby sources, and exhibit weak scintillation (ie. with $\sigma_I / \langle I \rangle \ll 1$, see J1136+1551 in Fig \ref{fig:spectraexamples}). The fact that the nearest pulsars show multiple screens suggests that there are many screens along any given line of sight.  However, strongly scintillating sources tend to show a single, dominant arc (e.g. J0922+0638 in Fig. \ref{fig:spectraexamples}).  This can arise either from physical reasons or from selection -  it is possible that there tends to be one dominant screen along the line-of-sight, perhaps due to having the highest $\Delta$DM, and being near to the halfway point which maximises the path length difference (and thus $\tau$) for a given deflection angle.  In addition, 2-screen effects could quench the scintillation from additional screens once the source appears extended, while in weakly scattered sources the pulsar line-of-sight image dominates and 2-screen effects are subdominant.  A possible selection effect is that sources of strong scattering result in inverted arclets, which may drown out additional screens in the secondary spectra.  One additional selection effect to consider is that all of the multi-screen sources have large proper motions of $140-660$\,km/s, where the slowest of these sources (J1057$-$5226) had a comparatively long 87.8\,min observation.  This suggests that the detection of multiple arcs is greatly influenced by the resolution in $f_D$, in addition to the effects described above.  We discuss future avenues to break these ambiguities in Section \ref{sec:discussion}.

The outermost arcs in three of our multi-arc sources imply scattering screens very near the source, with $d_{pl} \approx 35\,$pc, $d_{pl} \approx 27\,$pc, $d_{pl} \approx  65\,$pc in J0536$-$7543, J1239+2453, and J2048$-$1616 respectively, or within a parsec of the Earth. 
If the scattering is near Earth, there would be a strong annual modulation of the arcs, similar to the outermost arc in B1133+16 \citep{mckee+22}, and the corresponding screens would have a large angular extent.  These multi-screen sources imply a screen density along the line of sight of $\lesssim 200-300\,$pc/screen, while the 6 screens in B1133+16 shown by \citet{mckee+22} imply $\lesssim 60\,$pc/screen.  

\subsection{Secondary Spectrum Evolution with Frequency}
\label{sec:evolution}

In Figure \ref{fig:SecspecMultiscreen}, in addition to multi-screen sources, we included sources which show interesting behaviour across frequency.  This includes sources with discrete features at the same location across frequency, consistent with scattering from compact regions of fixed angular position.  PSR J1614+0737 shows an evolution from weak to strong scattering across the band, showing evidence of inverted arclets at 1\,GHz indicative of a highly anisotropic screen.  PSR J1740$-$1311 looks like a diffuse arc at $1\,$GHz, but appears to show structured power at $1.4\,$GHz.  A quantitative comparison of the feature positions across frequency (e.g. \citealt{brisken+10, rickett+21}), and with time (e.g. \citealt{hill+05, sprenger+22}) will concretely determine if scattering occurs from compact regions, and may distinguish between scattering from over- or under-dense regions.

\subsection{PSR J1731--4744}
\label{sec:J1744}
\begin{figure}
\centering
\includegraphics[trim=0.0cm 0cm 0cm 0.8cm, clip=true, width=1.0\columnwidth]{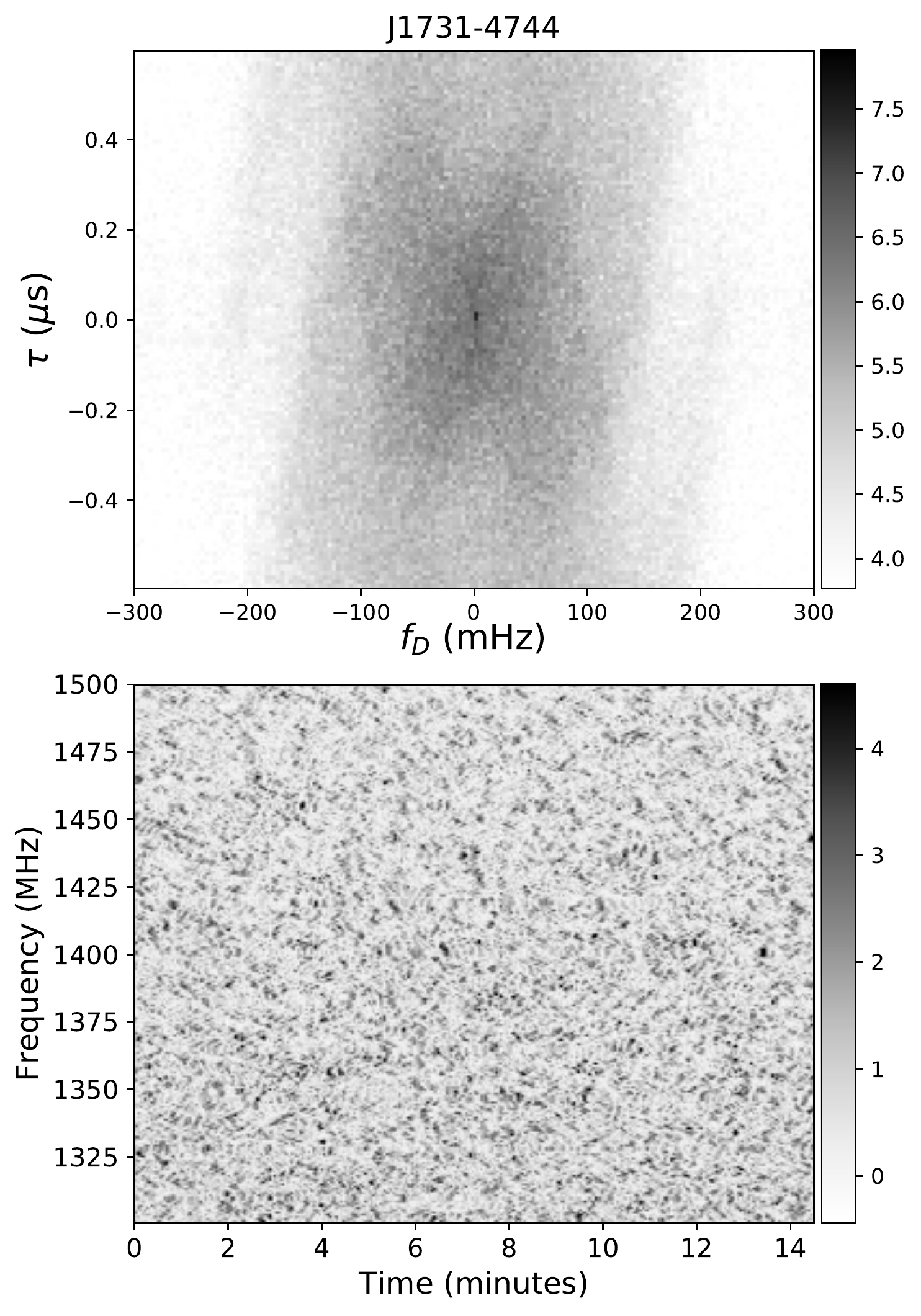}
\vspace{-5mm}
\caption{Dynamic and secondary spectrum of J1731$-$4744, rereduced with single pulses as described in section \ref{sec:J1744}. The large effective velocity implies a screen distance of $d_{\rm ls}\sim12\,$pc, comparable to the radius of SNR RCW114 which the pulsar is coincident with.}
\label{fig:J1731-4744}
\end{figure}

PSR J1731$-$4744 (B1727$-$47) was recently determined to be a high proper-motion pulsar with $\mu=151\pm19$\,mas/year, and tracing back the pulsar motion suggests an association with supernova remnant RCW 114 \citep{shternin+19}.  In our sample, J1731$-$4744 stood out as having the highest value of $W$, or equivalently showing an extremely fast $\sim$several second scintillation timescale with resolvable scintles in frequency, shown in Fig. \ref{fig:J1731-4744}. For this source, we re-reduced the filterbank data with single pulse time integrations ($\approx0.83$\,s), to fully resolve the arc in $f_D$.  The measured value of $W=3830\pm440$\,\zu indicates either a screen near to the source, with estimates of $s\approx0.02$, and $d_{pl}\approx12$\,pc, or a screen at $d_l \approx 0.03\,$pc from the Earth.  
The SNR shell of RCW 114 has a radius of $\approx2^{\circ}$, $\approx 20\,$pc at $d_{\rm p}\approx600$\,pc, comparable to $d_{\rm pl}$.  Our result strongly suggests the association of J1731$-$4744 with RCW 114, in agreement with \citet{shternin+19}.

\subsection{Comparison with Literature}

Several of the sources in our sample have had previously observed scintillation arcs.  We highlight and compare a few of these results.

\paragraph*{J0837+0610/B0834+06} is one of the best studied pulsars for scintillation, showing dramatic inverted arclets, being used for phase retrieval \citep{walker+08}, and imaged through VLBI to be a highly anisotropic screen \citep{brisken+10} at a distance of $d_{l} \approx 420$\,pc. In our data, we detect only a forward arc without inverted arclets.  Our estimate of $d_{l}\approx 412\,$pc is quite close to the more robust value from \citet{brisken+10}, suggesting we are still seeing the same scattering screen, where this agreement is likely aided by the fact that the screen is (seemingly coincidentally) closely aligned with the pulsar proper motion. The transition from strong to weak scattering is consistent with the findings of \citet{smirnova+20}, who observed the same change in the source's scintillation between 2012 and 2015.

\paragraph*{J1057-5226/B1055-52} was studied with Parkes data at 1400\,MHz by \citet{kerr+18}, finding a prolonged period of increased scattering (ie. decreased scintillation bandwidth) from MJD 54725--55725, interpreted as an extreme scattering event (ESE).  During the ESE, the secondary spectra were dominated by a single arc with inverted arclets at curvature of $\eta_{\rm outer}\approx0.02$s$^3$.  In observations at MJD 57506--57509, long past the ESE, a second, interior arc is visible at $\eta_{\rm inner}\approx0.06$s$^3$.  These arcs correspond to $W_{\rm outer} \approx 340$\,\zu, $W_{\rm inner} \approx 200$\,\zu, in close agreement with the two arcs seen in our data at $W_{\rm outer} = 372\pm79$\,\zu,  $W_{\rm inner} = 166\pm14$\,\zu on MJD 58655.  This suggests that the nature of scattering is similar to the state of `quiescent' scattering observed in the later observations of \citet{kerr+18}; the same arcs being visible over many years, combined with the pulsar's transverse velocity of $\approx 140$\,km/s implies the transverse extent of the two screens is $\gtrsim 100$\,AU, $\gtrsim 300$\,AU respectively.

\paragraph*{J1136+1551/B1133+16} was studied by \citet{mckee+22}, who measure the variations of 6 arcs across 40 years.  In our single observation, we clearly detect three of these arcs, which correspond to their arcs B, C, and E. In longer observations with finer frequency resolution, we could likely further disentangle the multiple screens.

\paragraph*{J1239+2453/B1237+25} shows clear evidence of 3 screens in our data, across the full band.  In previous LOFAR observations at 110-190\,MHz, the secondary spectra show only a diffuse blend of power \citep{wu+22}.  Additionally, only the inner arc was faintly seen by \citet{fadeev+18}, who observed the source with GBT at $316-332$\,MHz.

\paragraph*{J1921+2153/B1919+21} showed a clear arc at LOFAR in \citet{wu+22}. Their values of $\eta\approx 4.4\,s^3$ at observing frequency of $150\,$MHz correspond to $W\approx216$\,\zu, in agreement with our measured value of $W=243\pm25$\,\zu, suggesting we are observing the same screen.

\paragraph*{J0452$-$1759/B0450$-$18} was studied in detail by \citet{rickett+21} at 340 and 800\,MHz with the GBT, showing clear inverted arclets and being fit well with a 1D screen brightness distribution.  In our data, we miss this structure, seeing only the outline of an arc.  This shows the limitations of our short observing time (9\,min, compared to 60\,min observations), and that rich structures can be revealed using longer integrations.  It is also possible that our observation is at a low point of $v_{\rm eff,||}$ from annual motion.

\section{Ramifications}
\label{sec:discussion}

In summary, we have presented the discovery of 107 sources with detectable scintillation arcs with the Meerkat TPA.  In this section we discuss ramifications and future avenues.

Our findings are consistent with recent suggestions that scintillation arcs are prevalent \citep{stinebring+22}, 
their detection primarily limited by frequency resolution and observation duration, provided the dynamic spectrum is high S/N and largely devoid of RFI.
The nature of this survey manifests in strong biases, where arcs are easiest to detect in weakly scattered sources with high effective velocity - either through a high proper motion, or local screens, as in the case of J1731$-$4744. The detection of multi-arc sources include even more selection biases, with multi-arc sources primarily being nearby, weakly scintillating sources with very high transverse velocity.  There are likely many screens on any given sightline which cannot be seen in secondary spectra due to insufficient time duration, 2-screen interactions, or inverted arcs.

This survey leaves many avenues for further research.  In the TPA alone, there are hundreds more sources with resolvable scintillation, but without the detection of arcs. Analysis of the full sample can provide measures of scattering time $\tau$ on many more sightlines, which will help improve galactic electron models (e.g. \citealt{cordes+02, yao+17}).  The more distant, scattered sources will be best studied at higher frequencies where arcs with be sharper, more easily resolvable.  
Through wide-band/multi-frequency observations, or through phase-retrieval techniques such as cyclic spectroscopy \citep{demorest11, walker+13}, one could measure the correspondence between time-domain scattering and scintillation arcs. This may help elucidate the morphology of scattering tails, and investigate if seemingly isotropic scattering could be reproduced through the ensemble of many anisotropic screens (e.g. \citealt{oswald+21}).

With only a single arc curvature measurement, only crude estimates of screen distances can be made.  Additional observations to obtain annual curves \citep{reardon+20, main+20, mckee+22, mall+22}, or VLBI \citep{brisken+10} can fully solve for screen distances and geometry. In multi-screen sources, VLBI combined with dynamic cross-spectra can be used to simultaneously solve for multiple screens \citep{simard+19b}, and characteristic 2-screen effects may also provide additional constraints on screen geometry, as seen in B1508+55 \citep{sprenger+22}.  With a growing number of scattering screen measurements, it may be possible to map the distribution of screens in the Milky Way, cross-matching with foreground stars, HII regions, supernova remnants, and with the local bubble and other local plasma.
With high S/N, one could probe magnetism in scattering screens, and compare scintillation across pulse phase to resolve pulsar emission.  The results of this survey will then serve as a valuable stepping stone to studying the distribution of free electrons in the Milky Way, probing the dynamics and emission of pulsars, and elucidating the astrophysical origin of interstellar scintillation.

\section*{Acknowledgements}
We thank the reviewer William Coles for many useful comments which improved this manuscript.
RAM thanks Tim Sprenger and Olaf Wucknitz for their discussions and insights, and Fang Xi Lin and Daniel Baker for help with the Wiener Filter. LSO acknowledges the support of Magdalen College, Oxford. 
The MeerKAT telescope is operated by SARAO, which is a facility of the National Research Foundation, an agency of the Department of Science and Innovation.
MeerTime data are housed and processed on the OzSTAR supercomputer at Swinburne University of Technology, funded
by Swinburne University of Technology and the National Collaborative Research Infrastructure Strategy (NCRIS). 

\section*{Data Availability}
The dynamic spectra used in this paper are included on Zenodo at \url{https://doi.org/10.5281/zenodo.7261413}.

\bibliography{main}{}
\bibliographystyle{mnras}

\appendix
\renewcommand{\thefigure}{A\arabic{figure}}
\setcounter{figure}{0}
\clearpage

\begin{table}
\begin{center}
\caption{Arc measurements in sources with known proper motions. $W$ and $\Delta W_{lr}$ are computed as described in Section \ref{sec:arcfitting}. Values of d$_{\rm psr}$ with errorbars come from parallax measurements cited below, otherwise they are DM distances from NE2001 \protect{\citep{cordes+02}}.  In $\Delta W_{lr}$, $-$ or $+$ denote cases where the arc could only be measured on the negative or positive side of $f_{\rm D, norm}$. Distance references: (1) \citet{deller+09}, (2) \citet{brisken+03}, (3) \citet{chaterjee+09}, (4) \citet{liu+16}, (5) \citet{chaterjee+01}, (6) \citet{deller+19}, (7) \citet{brisken+01}    }
\begin{tabular}{ cccccc }
 Source & DM & $d_p$ & V$_{\rm pm}$ & $W$ & $\Delta W_{lr}$  \\
              & pc cm$^{-3}$ & kpc & km s$^{-1}$ & km s$^{-1}$ & km s$^{-1}$ \\
              &  &  &  & kpc$^{-0.5}$ & kpc$^{-0.5}$ \\
\hline
J0151-0635 & 25.66 & 1.2 & 69 & $88\pm29$ & $-$  \\
J0206-4028 & 12.9 & 0.6 & 76 & $98\pm36$ & $-$  \\
J0304+1932 & 15.657 & 0.6 & 110 & $51\pm3$ & $-4\pm3$  \\
J0452-1759 & 39.903 & 2.4 & 145 & $134\pm25$ & $-35\pm13$  \\
J0536-7543 & 18.58 & 0.8 & 255 & $1488\pm18$ & $-25\pm17$  \\
.. & .. & .. & .. & $907\pm83$ & $-117\pm7$ \\
.. & .. & .. & .. & $574\pm28$ & $-$  \\
.. & .. & .. & .. & $191\pm119$ & $-$  \\
J0630-2834$^1$ & 34.425 & 0.333$^{+0.051}_{-0.039}$ & 81 & $183\pm38$ & $-54\pm14$  \\
J0659+1414$^2$ & 13.94 & 0.288$^{+0.033}_{-0.027}$ & 60 & $209\pm84$ & $-118\pm33$  \\
J0758-1528 & 63.327 & 3.0 & 219 & $127\pm38$ & $54\pm28$  \\
J0820-1350$^3$ & 40.938 & 1.961$^{+0.167}_{-0.109}$ & 418 & $322\pm57$ & $-27\pm57$  \\
J0837+0610$^4$ & 12.864 & 0.613$^{+0.062}_{-0.052}$ & 148 & $267\pm6$ & $-7\pm6$  \\
J0908-1739 & 15.879 & 0.9 & 208 & $345\pm47$ & $63\pm47$  \\
J0922+0638$^5$ & 27.299 & 1.205$^{+0.224}_{-0.163}$ & 505 & $338\pm64$ & $91\pm9$  \\
J1057-5226 & 29.69 & 0.7 & 143 & $166\pm14$ & $-20\pm6$  \\
.. & .. & .. & .. & $372\pm79$ & $50\pm80$  \\
J1136+1551$^6$ & 4.841 & 0.372$^{+0.002}_{-0.002}$ & 660 & $1171\pm226$ & $-31\pm229$  \\
.. & .. & .. & .. & $764\pm121$ & $26\pm122$  \\
.. & .. & .. & .. & $419\pm20$ & $-$  \\
J1141-6545 & 116.08 & 2.5 & 40 & $193\pm36$ & $47\pm38$  \\
J1239+2453$^7$ & 9.252 & 0.862$^{+0.064}_{-0.056}$ & 472 & $2462\pm152$ & $-318\pm152$  \\
.. & .. & .. & .. & $1288\pm59$ & $66\pm59$  \\
.. & .. & .. & .. & $342\pm22$ & $15\pm6$  \\
J1257-1027 & 29.634 & 1.6 & 103 & $306\pm35$ & $50\pm13$  \\
J1430-6623 & 65.102 & 1.0 & 176 & $305\pm40$ & $-57\pm24$  \\
J1534-5334 & 24.82 & 1.1 & 160 & $119\pm67$ & $95\pm17$  \\
J1645-0317$^6$ & 35.756 & 3.968$^{+0.324}_{-0.397}$ & 386 & $287\pm63$ & $-74\pm63$  \\
J1650-1654 & 43.25 & 1.5 & 114 & $146\pm28$ & $39\pm15$  \\
J1703-1846$^6$ & 49.551 & 2.857$^{+0.476}_{-0.357}$ & 229 & $267\pm38$ & $54\pm12$  \\
J1731-4744 & 123.056 & 0.6 & 429 & $3826\pm440$ & $-489\pm443$  \\
J1740+1311 & 48.668 & 1.5 & 208 & $365\pm15$ & $13\pm11$  \\
J1825-0935 & 19.383 & 0.9 & 66 & $82\pm22$ & $-31\pm16$  \\
J1844+1454 & 41.486 & 2.2 & 476 & $126\pm59$ & $-$  \\
J1900-2600 & 37.994 & 1.2 & 286 & $415\pm64$ & $48\pm64$  \\
J1901-0906$^6$ & 72.677 & 1.961$^{+0.167}_{-0.237}$ & 183 & $189\pm36$ & $-50\pm26$  \\
J1921+2153 & 12.444 & 1.1 & 186 & $243\pm25$ & $-32\pm25$  \\
J1941-2602 & 50.036 & 1.7 & 123 & $131\pm34$ & $-47\pm34$  \\
J1946-2913 & 44.309 & 1.5 & 277 & $76\pm8$ & $11\pm8$  \\
J2046-0421 & 35.799 & 1.7 & 96 & $50\pm44$ & $-63\pm22$  \\
J2048-1616$^3$ & 11.456 & 0.952$^{+0.018}_{-0.026}$ & 511 & $466\pm43$ & $-66\pm21$  \\
.. & .. & .. & .. & $1508\pm35$ & $-78\pm17$ \\
.. & .. & .. & .. & $1927\pm109$ & $-225\pm109$ \\
J2144-3933$^1$ & 3.35 & 0.165$^{+0.017}_{-0.014}$ & 130 & $319\pm36$ & $-51\pm5$  \\
J2317+2149$^6$ & 20.87 & 1.961$^{+0.213}_{-0.206}$ & 79 & $56\pm14$ & $-$  \\
J2330-2005 & 8.456 & 0.4 & 138 & $143\pm8$ & $-11\pm5$  \\
J2346-0609$^6$ & 22.504 & 3.636$^{+0.548}_{-0.258}$ & 733 & $534\pm23$ & $+$  \\
\hline
\label{table:arcmeas}
\end{tabular}
\footnotesize{$^{*}$:\, J1141$-$6545 is dominated by its binary motion, resulting in orbital variations of $W$ \protect{\citep{reardon+19}}. }
\end{center}
\end{table}
\begin{table}
\begin{center}
\caption{Arc measurements and estimates of $V_{\rm ISS}$ in sources without known proper motions. The errors on $V_{\rm ISS}$ include only the measurement error on $W$, not the large unknown uncertainty in the distance, and are subject to the assumptions and biases discussed in section \ref{sec:approx}.}
\begin{tabular}{ cccccc }
 Source & DM & d$_p$ & $W$ & $V_{\rm ISS}$ & $\Delta W_{lr}$ \\
              & pc cm$^{-3}$ & kpc & km s$^{-1}$ & km s$^{-1}$ & km s$^{-1}$ \\
              &  &  &  kpc$^{-0.5}$ & & kpc$^{-0.5}$ \\
\hline
J0211-8159 & 24.36 & 1.0 & $224\pm35$ & $224\pm35$ & $9\pm35$ \\
J0343-3000 & 20.2 & 0.9 & $170\pm8$ & $164\pm8$ & $-$ \\
J0455-6951 & 94.7 & 43.4 & $23\pm9$ & $152\pm59$ & $+$ \\
J0459-0210 & 21.02 & 0.9 & $144\pm17$ & $138\pm16$ & $-7\pm17$ \\
J0529-6652 & 103.31 & 46.2 & $378\pm91$ & $2569\pm619$ & $+$ \\
J0543+2329 & 77.703 & 2.1 & $402\pm129$ & $577\pm185$ & $-182\pm48$ \\
J0633-2015 & 90.7 & 6.6 & $27\pm8$ & $69\pm21$ & $-$ \\
J0636-4549 & 26.31 & 1.3 & $376\pm78$ & $431\pm89$ & $-110\pm76$ \\
J0656-2228 & 32.39 & 1.9 & $117\pm14$ & $161\pm19$ & $9\pm14$ \\
J0909-7212 & 54.3 & 1.9 & $96\pm74$ & $132\pm102$ & $105\pm12$ \\
J0919-6040 & 82.5 & 2.5 & $34\pm12$ & $54\pm19$ & $-$ \\
J0924-5814 & 57.4 & 1.8 & $140\pm17$ & $190\pm23$ & $14\pm17$ \\
J1001-5939 & 113.0 & 2.8 & $457\pm38$ & $758\pm63$ & $-54\pm38$ \\
J1018-1642 & 48.82 & 3.2 & $89\pm45$ & $158\pm80$ & $-63\pm16$ \\
J1020-5921 & 80.0 & 2.1 & $154\pm28$ & $222\pm40$ & $21\pm28$ \\
J1032-5206 & 139.0 & 3.5 & $161\pm15$ & $299\pm28$ & $-$ \\
J1055-6905 & 142.8 & 3.9 & $118\pm51$ & $234\pm101$ & $31\pm51$ \\
J1112-6926 & 148.4 & 4.0 & $106\pm29$ & $212\pm58$ & $-41\pm19$ \\
J1119-7936 & 27.4 & 1.0 & $116\pm17$ & $117\pm17$ & $-1\pm17$ \\
J1132-4700 & 123.0 & 5.2 & $262\pm81$ & $596\pm184$ & $-114\pm48$ \\
J1141-3107 & 30.77 & 1.2 & $257\pm47$ & $283\pm52$ & $-$ \\
J1210-6550 & 37.0 & 1.1 & $125\pm38$ & $134\pm41$ & $-54\pm5$ \\
J1232-4742 & 26.0 & 0.9 & $129\pm12$ & $125\pm12$ & $4\pm15$ \\
J1239-6832 & 94.3 & 2.1 & $128\pm11$ & $186\pm16$ & $-2\pm11$ \\
J1240-4124 & 44.1 & 1.5 & $150\pm59$ & $185\pm73$ & $84\pm22$ \\
J1312-6400 & 93.0 & 1.9 & $78\pm8$ & $109\pm11$ & $+$ \\
J1340-6456 & 76.99 & 1.7 & $150\pm54$ & $196\pm70$ & $-$ \\
J1404+1159 & 18.499 & 1.4 & $97\pm4$ & $115\pm5$ & $12\pm3$ \\
J1418-3921 & 60.49 & 1.2 & $27\pm8$ & $29\pm9$ & $11\pm7$ \\
J1440-6344 & 124.2 & 3.1 & $141\pm28$ & $247\pm49$ & $-40\pm20$ \\
J1443-5122 & 87.0 & 1.9 & $94\pm42$ & $131\pm59$ & $60\pm11$ \\
J1514-4834 & 51.5 & 1.3 & $176\pm30$ & $201\pm34$ & $-38\pm30$ \\
J1527-3931 & 48.8 & 1.5 & $120\pm29$ & $147\pm36$ & $-$ \\
J1530-5327 & 49.6 & 1.2 & $1064\pm218$ & $1181\pm242$ & $308\pm59$ \\
J1535-4114 & 66.28 & 2.0 & $160\pm20$ & $224\pm28$ & $3\pm21$ \\
J1537-4912 & 69.7 & 2.5 & $107\pm20$ & $168\pm31$ & $-$ \\
J1549+2113 & 24.055 & 2.0 & $480\pm25$ & $673\pm35$ & $-36\pm17$ \\
J1603-2712 & 46.201 & 1.6 & $218\pm45$ & $272\pm56$ & $-64\pm25$ \\
J1605-5257 & 34.9 & 2.2 & $203\pm57$ & $301\pm84$ & $-80\pm26$ \\
J1614+0737 & 21.395 & 1.0 & $524\pm8$ & $515\pm8$ & $1\pm8$ \\
J1646-6831 & 43.0 & 1.3 & $23\pm7$ & $26\pm8$ & $0\pm7$ \\
J1720+2150 & 40.719 & 2.3 & $49\pm15$ & $74\pm23$ & $-74\pm8$ \\
J1720-1633 & 44.83 & 1.3 & $337\pm69$ & $390\pm80$ & $-98\pm11$ \\
J1728-0007 & 41.09 & 1.6 & $298\pm99$ & $374\pm124$ & $141\pm67$ \\
J1740+1000 & 23.897 & 1.2 & $456\pm191$ & $508\pm213$ & $271\pm26$ \\
J1745-0129 & 89.3 & 3.1 & $137\pm60$ & $242\pm106$ & $85\pm30$ \\
J1755-0903 & 63.67 & 1.8 & $257\pm23$ & $344\pm31$ & $-32\pm21$ \\
J1800-0125 & 51.0 & 1.8 & $641\pm37$ & $850\pm49$ & $40\pm41$ \\
J1810-5338 & 45.0 & 1.3 & $72\pm13$ & $82\pm15$ & $16\pm13$ \\
J1811-4930 & 44.0 & 1.2 & $118\pm6$ & $132\pm7$ & $+$ \\
J1813+1822 & 60.8 & 2.9 & $96\pm17$ & $164\pm29$ & $24\pm13$ \\
J1819+1305 & 64.808 & 2.8 & $107\pm16$ & $178\pm27$ & $22\pm12$ \\
J1825+0004 & 56.618 & 1.9 & $107\pm32$ & $148\pm44$ & $+$ \\
J1840-1419 & 19.4 & 0.8 & $26\pm5$ & $24\pm5$ & $7\pm3$ \\
J1849+0409 & 63.97 & 2.6 & $150\pm12$ & $241\pm19$ & $+$ \\
J1900-7951 & 39.0 & 1.4 & $179\pm40$ & $214\pm48$ & $57\pm13$ \\
J1917+0834 & 29.18 & 1.9 & $46\pm24$ & $64\pm33$ & $+$ \\
J1929+2121 & 66.0 & 3.4 & $42\pm7$ & $77\pm13$ & $-$ \\
J1943-1237 & 28.918 & 1.2 & $180\pm24$ & $198\pm26$ & $-33\pm28$ \\
\hline
\label{table:arcmeasnov}
\end{tabular}
\end{center}
\end{table}
\begin{table}
\begin{center}
\caption{Table 2 continued.}
\begin{tabular}{ cccccc }
 Source & DM & d$_p$ & $W$ & $V_{\rm ISS}$ & $\Delta W_{lr}$ \\
              & pc cm$^{-3}$ & kpc & km s$^{-1}$ & km s$^{-1}$ & km s$^{-1}$ \\
              &  &  &  kpc$^{-0.5}$ & & kpc$^{-0.5}$ \\
\hline
J1945-0040 & 59.71 & 2.5 & $76\pm8$ & $119\pm13$ & $0\pm8$ \\
J1949-2524 & 23.07 & 0.7 & $337\pm11$ & $281\pm9$ & $-$ \\
J1951+1123 & 31.29 & 2.1 & $100\pm5$ & $145\pm7$ & $1\pm5$ \\
J2005-0020 & 35.93 & 1.8 & $587\pm173$ & $781\pm230$ & $-$ \\
J2038-3816 & 33.96 & 1.4 & $163\pm43$ & $190\pm50$ & $-$ \\
J2043+2740 & 21.021 & 1.8 & $39\pm11$ & $52\pm15$ & $-15\pm3$ \\
J2108-3429 & 30.22 & 1.2 & $180\pm12$ & $198\pm13$ & $-$ \\
J2139+2242 & 44.16 & 2.8 & $101\pm14$ & $168\pm23$ & $-6\pm14$ \\
J2155-3118 & 14.85 & 0.6 & $55\pm12$ & $43\pm9$ & $-$ \\
J2253+1516 & 29.204 & 1.7 & $54\pm13$ & $70\pm17$ & $-$ \\
\hline
\end{tabular}
\end{center}
\end{table}

\begin{figure*}
\includegraphics[width=1.0\textwidth,trim=0.0cm 0.0cm 0cm 0.0cm, clip=true]{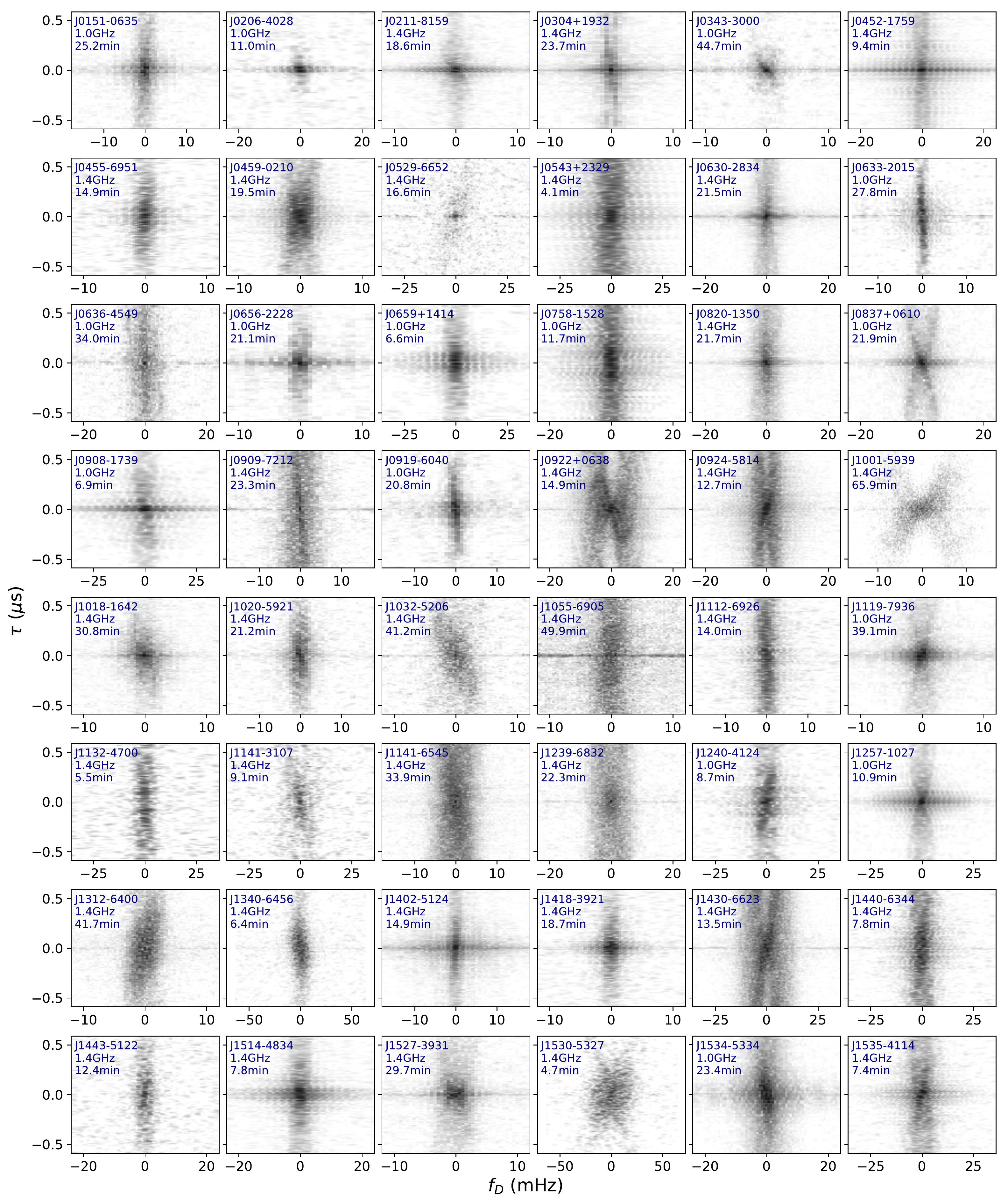} \\
\caption{Panorama of Secondary Spectra.  The colourbars are logarithmic, extending from the median to the maximum intensity.  The $\tau$ axes are all identical, covering the full observable range, while the $f_{D}$ range is chosen separately for each source to best show the arc.}
\label{fig:SecspecPano}
\end{figure*}   

\begin{figure*}
\includegraphics[width=1.0\textwidth,trim=0.0cm 0.0cm 0cm 0.0cm, clip=true]{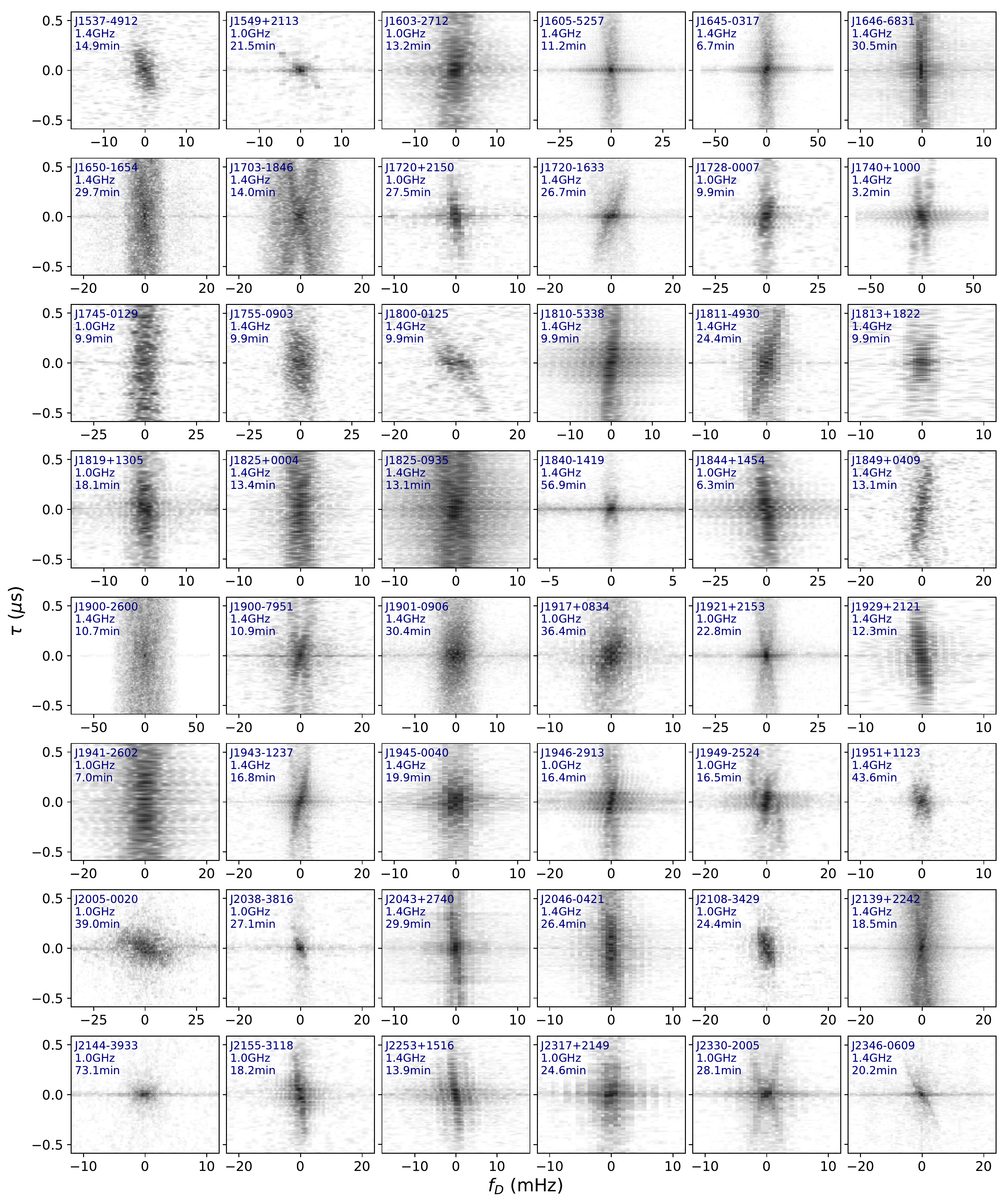} \\
\caption{Continued..}
\end{figure*}  

\bsp	
\label{lastpage}
\end{document}